\documentclass[galaxies,article,submit,moreauthors,pdftex,10pt,a4paper]{mdpi}

\firstpage{1}
\articlenumber{x}
\doinum{10.3390/------}
\pubvolume{xx}
\pubyear{2018}
\copyrightyear{2018}
\history{Received: date; Accepted: date; Published: date}
\newcommand{\wombat}{{\small WOMBAT }}
\usepackage{aas_macros}
\usepackage{amssymb}

\Title{Towards Exascale Simulations of the ICM Dynamo with WENO-Wombat}

\Author{Julius Donnert $^{1,2}$\orcidA{}, Hanbyul Jang$^{3}$, Peter Mendygral$^{4}$, Gianfranco Brunetti$^{1}$, Dongsu Ryu$^{3}$ and Thomas Jones$^{2}$}

\AuthorNames{Julius Donnert, Hanbyul Jang, Peter Mendygral, Gianfranco Brunetti, Dongsu Ryu, Thomas Jones}

\address{%
$^{1}$ \quad INAF Istituto di Radioastronomia, via P. Gobetti 101, I-40129 Bologna, Italy \\
$^{2}$ \quad School of Physics and Astronomy, University of Minnesota, Minneapolis, MN 55455, USA \\
$^{3}$ \quad Department of Physics, School of Natural Sciences, Ulsan National Institute of Science and Technology, UNIST-gil 50, Ulsan, 44919, Korea \\
$^{4}$ \quad Cray Inc., Bloomington, MN, USA
}
\corres{Correspondence: donnert@ira.inaf.it; Tel.: +39-051-6399364  }

\abstract{In galaxy clusters, modern radio interferometers observe non-thermal radio sources with unprecedented spatial and spectral resolution. For the first time, the new data allows to infer the structure of the intra-cluster magnetic fields on small scales via Faraday tomography.  This leap forward demands new numerical models for the amplification of magnetic fields in cosmic structure formation - the cosmological magnetic dynamo. Here we present a novel numerical approach to astrophyiscal MHD simulations aimed to resolve this small-scale dynamo in future cosmological simulations. As a first step, we implement a fifth order WENO scheme in the new code \wombat. We show that this scheme doubles the effective resolution of the simulation and is thus less expensive than common second order schemes. \wombat uses a novel approach to parallelization and load balancing developed in collaboration with performance engineers at Cray Inc. This will allow  us scale simulation to the exaflop regime and achieve kpc resolution in future cosmological simulations of galaxy clusters. Here we demonstrate the excellent scaling properties of the code and argue that resolved simulations of the cosmological small scale dynamo within the whole virial radius are possible in the next years.}


\begin{document}
\section{Introduction}

Most of the Baryonic matter in our Universe is in the form of magnetized plasma. Hence, astronomers observe the signature of astrophysical magnetic fields from the solar system to the large scale structure. In galaxy clusters, radio telescopes detect the synchrotron radiation ($50 \,\mathrm{MHz} - 30 \,\mathrm{GHz}$) emitted by relativistic electrons ($\gamma > 1000$) gyrating in the magnetic field ($B \sim 1 \,\mu\mathrm{G}$) of the intra-cluster-medium (ICM), a hot and underdense plasma ($T\sim10^8\,\mathrm{K}, n_\mathrm{th} \sim 10^{-3}\,\mathrm{cm}^{-3}$). The next generation of radio interferometers will infer the three dimensional structure of the field through \emph{Faraday tomography} on kpc scales. This represents a first serious probe of the small scale properties of the whole intra-cluster-medium that demands detailed predictions to interpret the new data. As radio brightness is not strongly correlated with thermal density, upcoming studies will probe the whole virial volume of a cluster. \par
The ICM itself is a weakly-collisional plasma, whose micro-physical properties are set by turbulence and electromagnetic interactions (plasma-waves), not particle Coulomb scattering \citep{2007mhet.book...85S}. Thus the magnetic field plays a crucial role in making the medium ''collisional'' on large scales, i.e. behave like a magnetised fluid \citep{2007MNRAS.378..245B}. In the currently favoured model of the ICM, the evolution of the magnetic field is governed by a turbulent small-scale dynamo that grows small seed fields at high redshift ($B \sim 10^{-13} \,\mathrm{G}$) into $\mu$G fields via an inverse cascade at the Alfv\'{e}n scale of the medium \citep{sch05,po15}. In merging galaxy clusters the Alfv\'{e}n scale\footnote{The scale where magnetic and turbulent pressure are comparable, i.e. where the Lorentz force becomes important \citep{1995ApJ...438..763G}} reaches a few kpc, thus it is now in range of next-generation radio interferometers for Faraday tomography studies. \par
The strength and geometry of the magnetic field is set by the local seeding and turbulence history of the gas parcel under consideration, thus the new data demands numerical simulations to compare with expectations from dynamo theory. However, the nature of the small scale dynamo has made it very difficult to obtain accurate numerical models for the ICM magnetic field \citep{bm16}. The crucial time scale of magnetic field growth is set by the smallest length scale available in the turbulent system, where the eddy turnover time is smallest. In nature, this can be far below pc scale, in simulations this is at best the resolution scale. Current state-of-the-art Eulerian simulations start seeing numerical effects below scales of 10 kpc, Lagrangian simulations reach better resolution in the cluster center, but do not come close at the cluster outskirts due to density adaptivity (see Donnert et al., subm. to SSRv for a review). Thus resolving the Alfv\'{e}n scale at ~3 kpc in the whole cluster volume and faithfully evolving the magnetic field through structure formation is not possible with current community codes. \par
In the preferred Eulerian approach, such simulations would require $\sim4096^3$ zones inside the virial radius, run with a highly accurate finite volume or finite difference scheme. This translates into 50--100 TBytes of memory and would generate 1--4 PByte of data. This is well in range of current Petascale and upcoming Exascale supercomputers, but requires near ideal weak scaling of the simulation code to 5--10 thousand compute nodes. Current state-of-the-art simulations typically run on a few thousand nodes, to maximize parallel efficiency \citep{2014ApJ...782...21M}. Hence, it stands to reason that in practice resolutions close the Alfv\'{e}n scale in the ICM will be challenging to achieve with current codes. \par
Here we present a performance-aware implementation of the a fifth order constrained transport weighted essentially non-oscillatory (WENO) scheme in the scalable \wombat code\footnote{\url{wombatcode.org}}. This implementation represents a first step towards simulations of the small-scale dynamo in the ICM in a cosmological framework that resolve the Alfv\'{e}n scale. We will show that WENO doubles the effective resolution of the simulation, but is only a factor 10 more computationally expensive than commonly used 2nd order schemes at the same resolution. Hence it is a more efficent algorithm. \wombat itself is an on-going research effort of performance engineers at Cray Inc. to maximize computational efficiency on upcoming exascale systems \citep{2017ApJS..228...23M}. We will show that the new code indeed achieves excellent performance on large supercomputers.  

\section{WENO-Wombat}

\subsection{WENO Algorithm}

The Weighted Essentially Non Oscillatory schemes \citep{1994JCoPh.115..200L} are an improvement of ENO schemes presented in \citet{1987JCoPh..71..231H,1988JCoPh..77..439S}. ENO schemes chose one out of several stencils around a zone $i$ based on a mathematical smoothness criterion to avoid spurious oscillations close to flow discontinuities (shocks). WENO schemes combine a weighted average of the stencils, so that the scheme is high order away from shocks, but still avoids Gibbs phenoma adjacent to them \citep[see][ for a review]{doi:10.1137/070679065,2017LRCA....3....2B}. WENO-Wombat implements the classical scheme from \citet{Jiang_Shu__1996__WENO-scheme,1999JCoPh.150..561J}, which combines three stencils to achieve formal fifth order trunctation error in space. The algorithm uses a Roe-type Riemann solver to decouple the system of 8 partial differential equations into independent advection equations \citep{ROE1997250}. The WENO interpolated fluxes on the faces of zone $i$ in the decoupled system are given by
\begin{align}
    {F}^s_{i+\frac{1}{2}} &= \frac{1}{12} \left( -{F}^s_{i-1}+7{F}^s_{i}+7{F}^s_{i+1}-{F}^s_{i+2} \right) -\varphi_N\left( \Delta{F}^{s+}_{i-\frac{3}{2}},\Delta{F}^{s+}_{i-\frac{1}{2}},\Delta{F}^{s+}_{i+\frac{1}{2}},\Delta{F}^{s+}_{i+\frac{3}{2}} \right) \nonumber\\
    &+\varphi_N\left( \Delta{F}^{s-}_{i+\frac{5}{2}},\Delta{F}^{s-}_{i+\frac{3}{2}},\Delta{F}^{s-}_{i+\frac{1}{2}},\Delta{F}^{s-}_{i-\frac{1}{2}}\right), \label{eq.FWENO}
\end{align}
where $\Delta{F}^{s-}_i$ the flux on the cell boundaries obtained from a simple Lax-Friedrichs splitting. The WENO5 interpolant $\varphi_N(a,b,c,d)$ is defined as
\begin{align}
    \varphi_N(a,b,c,d) &= \frac{1}{3} \omega_0 \left(a-2b+c\right) + \frac{1}{6} \left( \omega_2 - \frac{1}{2} \right) \left( b - 2c + d \right).
\end{align}
The non-linear weights are given by:
\begin{align}
    \omega_0 &= \frac{\alpha_0}{\alpha_0 + \alpha_1 + \alpha_2}, & \omega_2 &= \frac{\alpha_2}{\alpha_0 + \alpha_1 + \alpha_2} \\
    \alpha_0 &= \frac{1}{\left(\epsilon + IC_0\right)^2}, & \alpha_1 &= \frac{6}{\left(\epsilon + IC_1\right)^2}, & \alpha_2 &= \frac{3}{\left(\epsilon + IC_2\right)^2},
\end{align}
with $\epsilon = 10^{-6}$ and
\begin{align}
    IS_0 &= 13(a-b)^2 + 3 (a-3b)^2,\\
    IS_1 &= 13(b-c)^2 + 3 (b+c)^2,\\
    IS_2 &= 13(c-d)^2 + 3 (3c-d)^2.
\end{align}
Time integration is realized with a fourth order four-stage Runge Kutta integrator. \par
It has been shown that the resulting scheme is only third order accurate close to critical points (extrema) of the flow. In real world applications this is actually beneficial, because the scheme becomes more robust than a full fifth order scheme like WENO-Z \citep{2008JCoPh.227.3191B} that would fall back to protection fluxes instead.\par
Magnetic fields are treated in a constrained transport staggered mesh approach following \citet{1998ApJ...509..244R}, which is formally only second order accurate. However, using the high order fluxes the scheme conserves magnetic energy density to fifth order (Jang et al. in prep.), which is of crucial importance for the small dynamo. For the complete description of the algorithm including eigenvectors we refer the reader to the full method papers Jang et al. in prep. and Donnert et al. in prep.

\subsection{Wombat Implementation}

\wombat is a hybrid parallel MPI/OpenMP code written in object oriented Fortran 2008 \citep{2017ApJS..228...23M}. It is developed in collaboration with the programming environment group at Cray Inc., a major manufacturer of super computers. \par
The code initially divides the computational domain $\Omega$ into rectangular sub-domains (\emph{domains}) residing each on a separate MPI rank. Each sub-domain is further divided into independent pieces of work, rectangular \emph{patches}, usually of size $18^D-32^D$ zones, where $D$ is the number of dimensions. Patches and domains are implemented using Fortran 2008 objects, which makes the code highly modular. Patches carry ghost zones, domains carry ghost domains, which overlap with neighbouring MPI ranks and facilitate communication. Their size is tunable. If a patch is exported into a ghost domain, it is communicated to the according MPI rank owning the domain. This facilitates load-balancing among ranks. Ghost/boundary zones of patches are overlapping zones between neighbouring patches that need to be communicated for the patch to be computed. Once a patch has received all its boundary zones, it can be computed (\emph{resolved}) independently for the rest of the world grid. Depending on the solver, this may happen many times per time step. \par
\wombat implements communication of ghost zones and domains using fully one-sided thread-asynchronous MPI-RMA with \mbox{MPI\_THREAD\_MULTIPLE}. The scheme is continuously improved and an active research topic for the exa-scale  by performance engineers are Cray Inc.  To minimize OpenMP overhead the code uses a single OpenMP parallel region, in which all threads perform work and communication independently and asynchronously from each other. This requires the MPI library to support OpenMP lock-free communication. Patches internal to a rank are resolved in memory and can be immediately computed. Patches with boundary zones overlapping with another rank are communicated: The boundary zones are copied (''packed'') into mailboxes by the neighbouring rank, communicated with \mbox{MPI\_Get}, unpacked on the local rank and eventually the patch is resolved. The status of the mailbox (empty, packed) is communicated with 8 Byte signals (the ''heartbeat''). The heartbeat signals are always issued and facilitate a weak form of synchronization among ranks. Every step of the communication (packing, heartbeat signal, unpacking, resolution) can be done by any thread on the rank at any time. Thus the scheme achieves computation/communication overlap at the thread level and can react to imbalance from work decomposition or network contention on the machine. \par
We have implemented the WENO scheme into the \wombat framework as a separate solver module. At the beginning, the state vector grid of the resolved patch is copied into thread local memory. In multiple dimensions, it is flattened into a one dimensional array to increase vector length. This introduces memory overhead of about 25 Megabytes per thread and rank for $18^3$ zone patches. The whole algorithm operates on single index arrays, thus all loops are SIMD vectorized by the compiler. This is necessary to achieve a significant fraction of peak double precision performance on modern CPUs and greatly simplifies future GPU ports of the algorithm. In multiple dimensions, the grid has to be swept along the y-direction and z-direction. We implement the sweeps by re-ordering the data arrays, which corresponds to rotations of the grid in three dimensions. The resulting fluxes are then rotated back into the original layout, so that final updates can be performed. At the end of a sub-step, the grid is saved into the multi-index arrays in global memory.\par
Per time step, the boundaries of 3 zones are communicated 8 times, i.e. the patches is passed 8 times until it is resolved. The WENO and CT scheme require one pass each per sub-step.

\section{Results}

\subsection{Fidelity}
\begin{figure}
\centering
\includegraphics[width=0.45\textwidth]{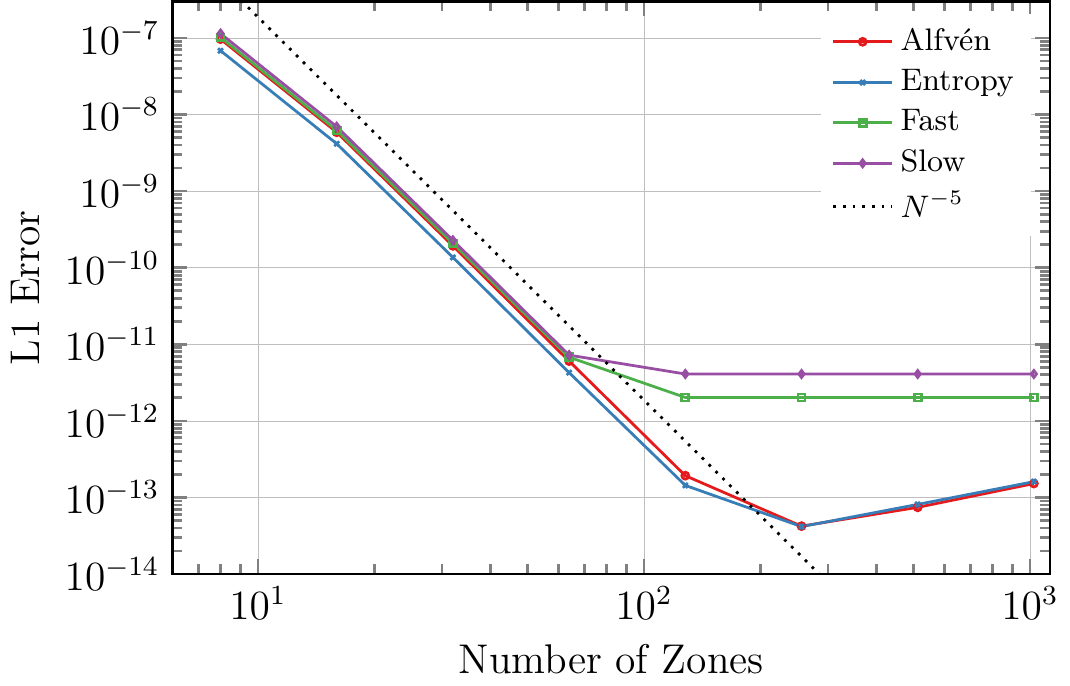}
\includegraphics[width=0.45\textwidth]{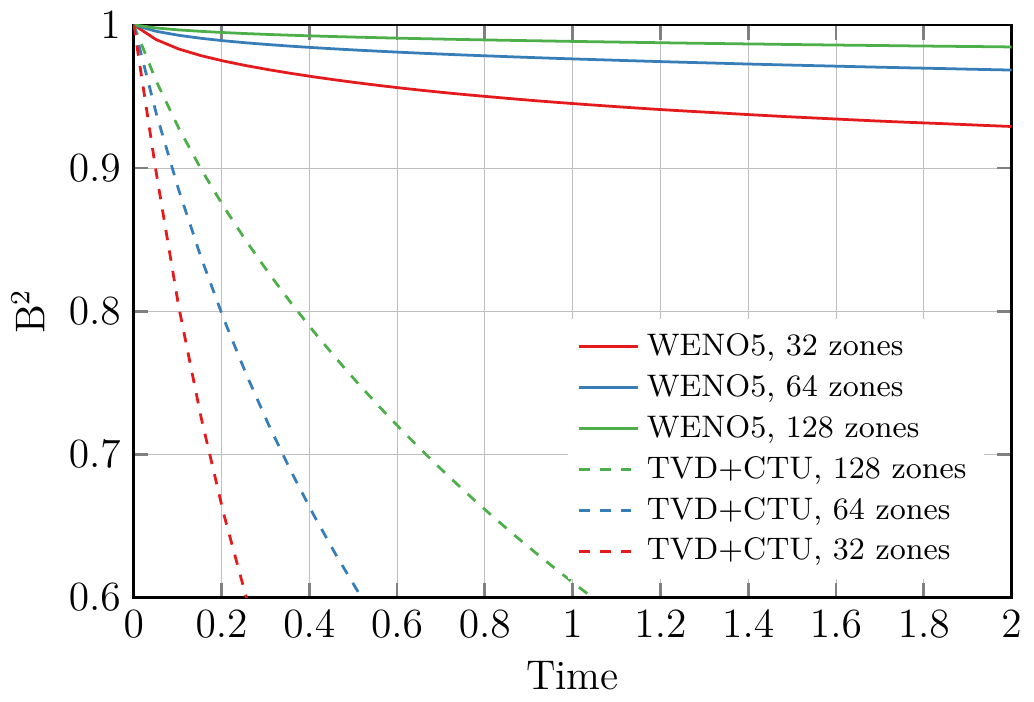}
\caption{Left: L1 error of the wave convergence test from in one dimension \citet{2008JCoPh.227.4123G}. Alfv\'{e}n mode in red, entropy mode in blue, fast mode in the green and slow mode in purple. Right: Evolution of magnetic energy in the advection of a field loop test.}
\label{fig.1d}
\end{figure}

We demonstrate the fidelity of the WENO MHD algorithm with a number of test cases. In figure \ref{fig.1d}, we show the convergence test involving advecting small ($10^{-5}$) perturbations in the four MHD waves across a one dimensional domain, following \citet{2008JCoPh.227.4123G}. All waves converge at fifth order down to L1 errors below $10^{-11}$. For the compressive fast and slow modes, wave steepening prevents further convergence in this test. Entropy and Alf\'{e}n mode converge to $10^{-13}$, where the roundoff error from 8 byte floating point precision prevents further convergence. A comparison with results from \texttt{ATHENA} \citep{2008JCoPh.227.4123G} shows that the WENO5 scheme improves on the CTU+CT
by more than a factor of two in effective resolution, i.e. \texttt{ATHENA} reaches L1 errors of $10^{-11}$ at 512 zones, WENO5 reaches theis L1 error at 64 zones. \par

\begin{figure}
\centering
\includegraphics[width=0.9\textwidth]{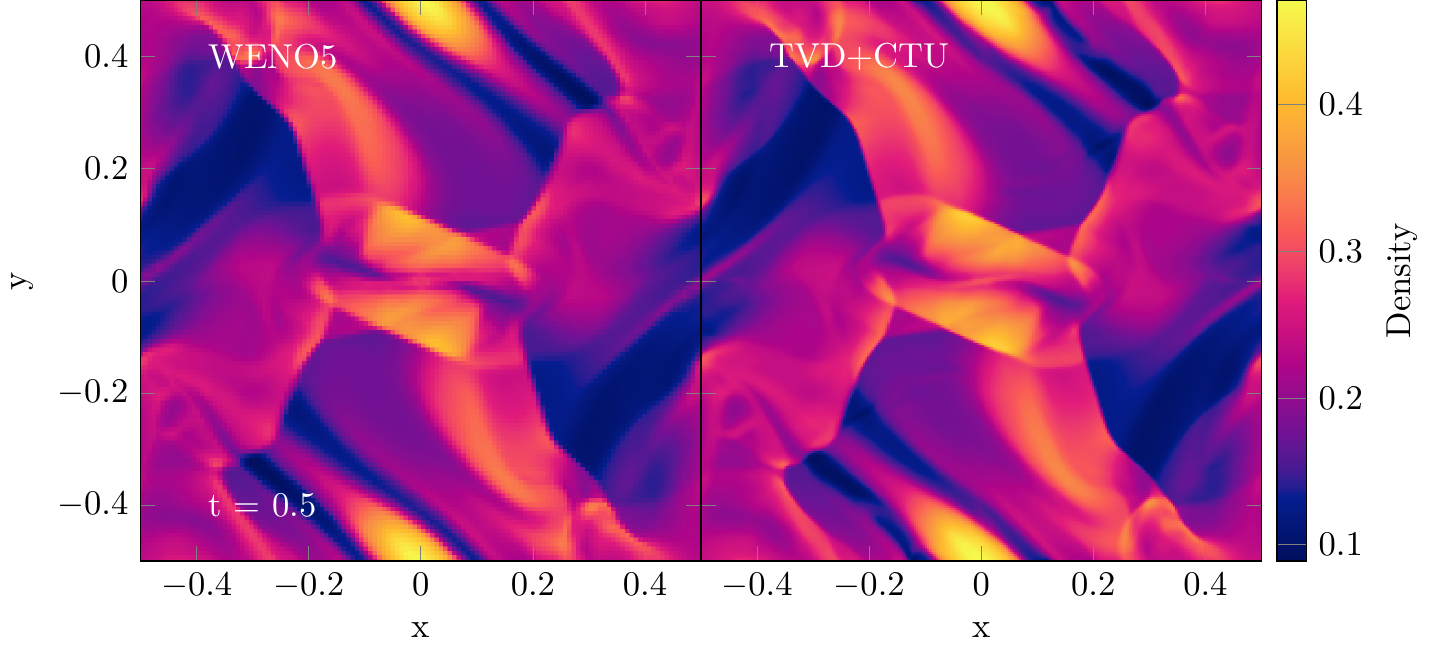} \\
\caption{Density in the 2D Orszag-Tang Vortex test. Left: WENO5 solution with $128^2$ zones. Right: TVD solution with $256^2$ zones.}
\label{fig.tests}
\end{figure}

\begin{figure}
\centering
\includegraphics[width=0.9\textwidth]{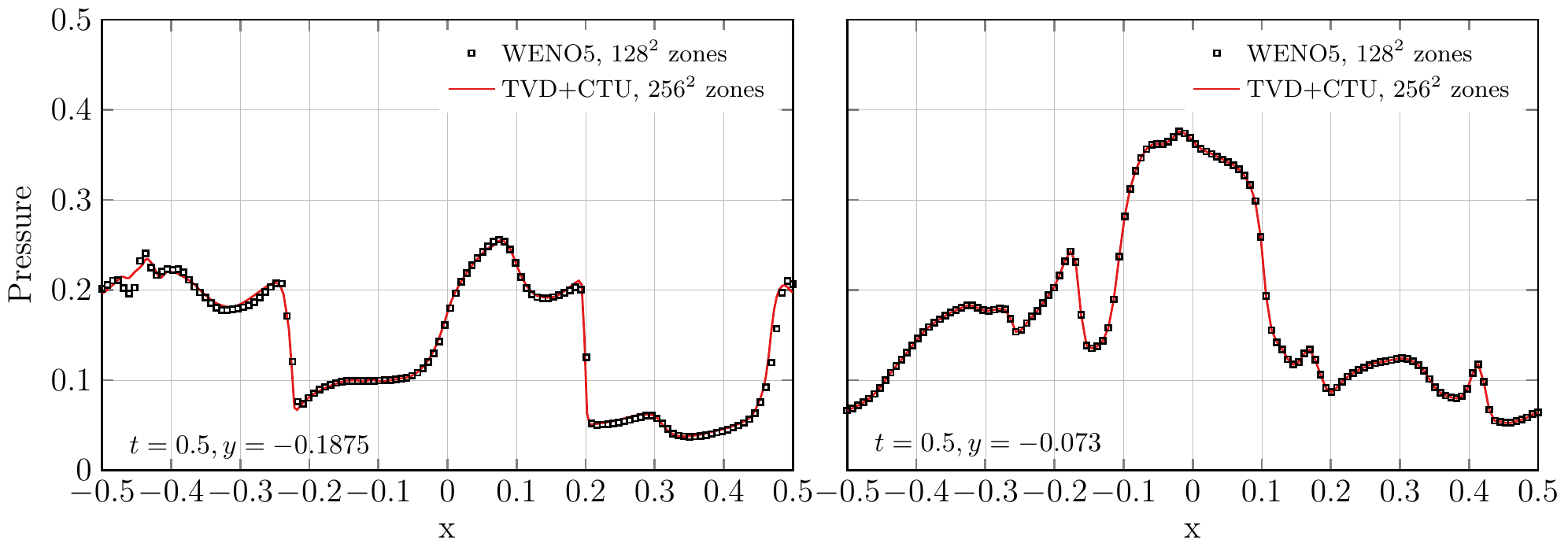}
\caption{Pressure slices in the 2D Orszag-Tang Vortex test at $y=-0.1875$ (left) and $y=-0.073$ (right). TVD solution with $256^2$ zones (red line), WENO5 solution with $128^2$ zones (black squares).}
\label{fig.tests3}
\end{figure}

In figure \ref{fig.tests} we show the density of the 2 dimensional Orszang-Tang vortex test from \citet{Orzang,2000JCoPh.161..605T,2008ApJS..178..137S}. The left panel shows the result from a run with \wombat's  new WENO5 implementation with $128^2$ zones. On the right the result from the run with 2nd order TVD+CTU implementation at $256^2$ zones. No difference is visible by eye. In figure \ref{fig.tests3} we show pressure slices of the test at $t=0.5$. Again WENO5 with $128^2$ zones resolves the complex pressure topology equally or better than the TVD+CTU result with $256^2$ zones. In particular, the pressure blip at $y=-0.1875, x=-0.45$. This demonstrates that the higher fidelity of the WENO5 algorithm effectively doubles the resolution of a MHD simulation. At the same time, a WENO5 run in 3D uses only an 8th of the memory and 75\% of the runtime of a TVD+CTU run at double its resolution. Thus we argue that fifth order WENO5 represents an \emph{optimal compromise between algorithmic fidelity and computational expense}. This result is in line with previous work on WENO3, WENO5 and WENO9: \citet{2003PhRvE..68d6709Z} found that every increase in order effectively doubles the resolution of the scheme. \par

\begin{figure}
\centering
\includegraphics[width=0.9\textwidth]{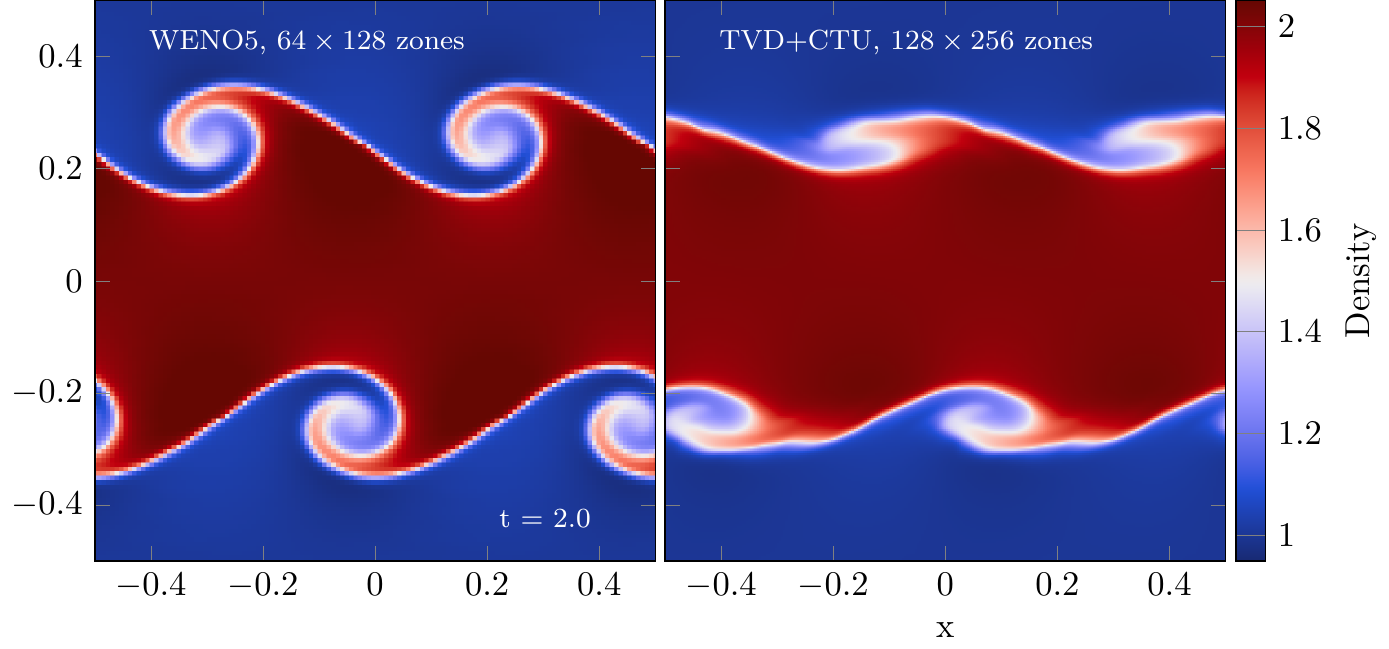}
\caption{Density in the 2D Kelvin Helmholtz test at time $t=2$. Left: WENO5 solution  with $64\times128$ zones. Right: TVD solution  with $128\times256$ zones. }
\label{fig.tests2}
\end{figure}
In figure \ref{fig.tests2}, we show a 2D Kelvin Helmholtz instability test following the ICs of \citet{2016MNRAS.455.4274L} who use a sinusoidal perturbation ($A = 0.01, P=10,a=0.05, \sigma=0.2, z_1=0.5,z_2=1.5, u_\mathrm{flow}=1, \Delta\rho = 1$) at $t=2$. We note that the with these parameters the instability is not resolved in both runs. \par
The left panel shows is the WENO5 solution with $64\times128$ zones. The right panel shows the TVD+CTU solution with twice the resolution. In the WENO5 simulation, the instability grows as expected and develops the famous vortices as well as fluctuations away from the shearing layer. In contrast, the second order run, does not show well developed vortices and substantially more diffusion. The growth of the instability is significantly slowed. This test shows that WENO5 resolves instabilities close to the resolution scale significantly better than a lower order code at twice the WENO5 resolution. \par
In figure \ref{fig.1d} right, we show the time evolution of magnetic energy in the field loop advection test from \citet{2008ApJS..178..137S}. Runs at $32^2, 64^2, 128^2$ zones in red, blue, green respectively are shown with WENO5 (solid lines) and with TVD+CTU (dashed lines). Magnetic energy is conserved much better in the WENO5 case, due to the fifth order fluxes in the scheme. A comparison with the \texttt{ATHENA} results presented in \citet{2008JCoPh.227.4123G} again show that WENO5 roughly doubles the effective resolution, e.g. at $t=1$ \texttt{ATHENA} with 128 zones conserves 95\% of magnetic energy (their figure 1), just as WENO5 with 64 zones. We argued above that magnetic field growth close to the resolution scale is the crucial mechanism for magnetic field amplification in galaxy clusters. Thus this result demonstrates the advantages of using a highly accurate CT scheme in cluster MHD simulations. \par

\subsection{Efficiency}

\begin{figure}
\centering
\includegraphics[width=0.49\textwidth]{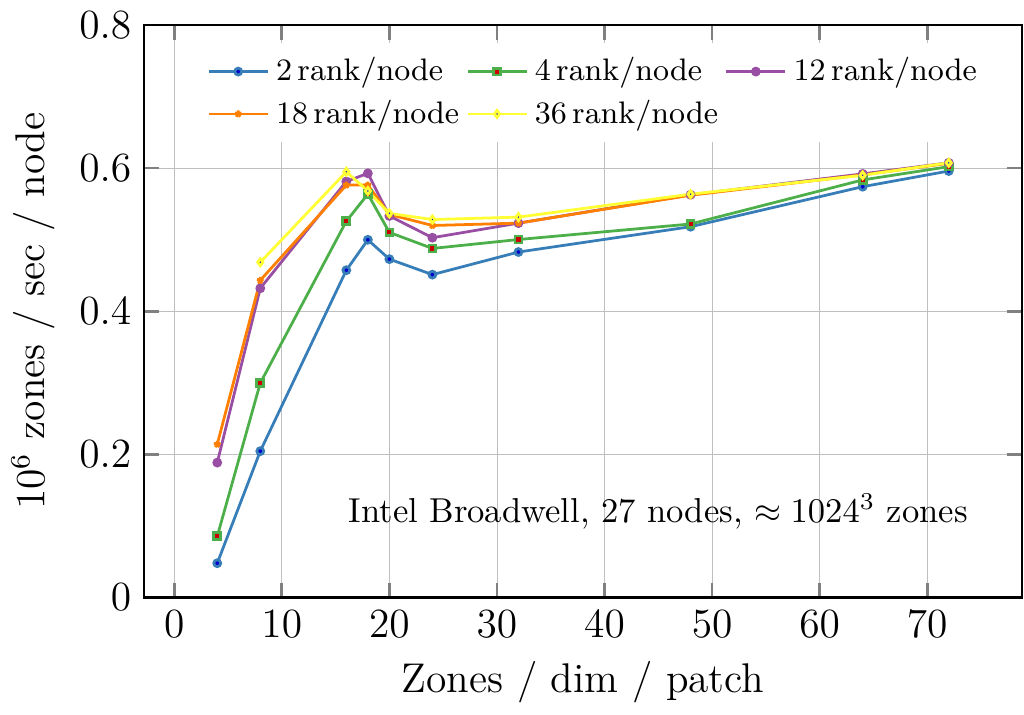}
\includegraphics[width=0.49\textwidth]{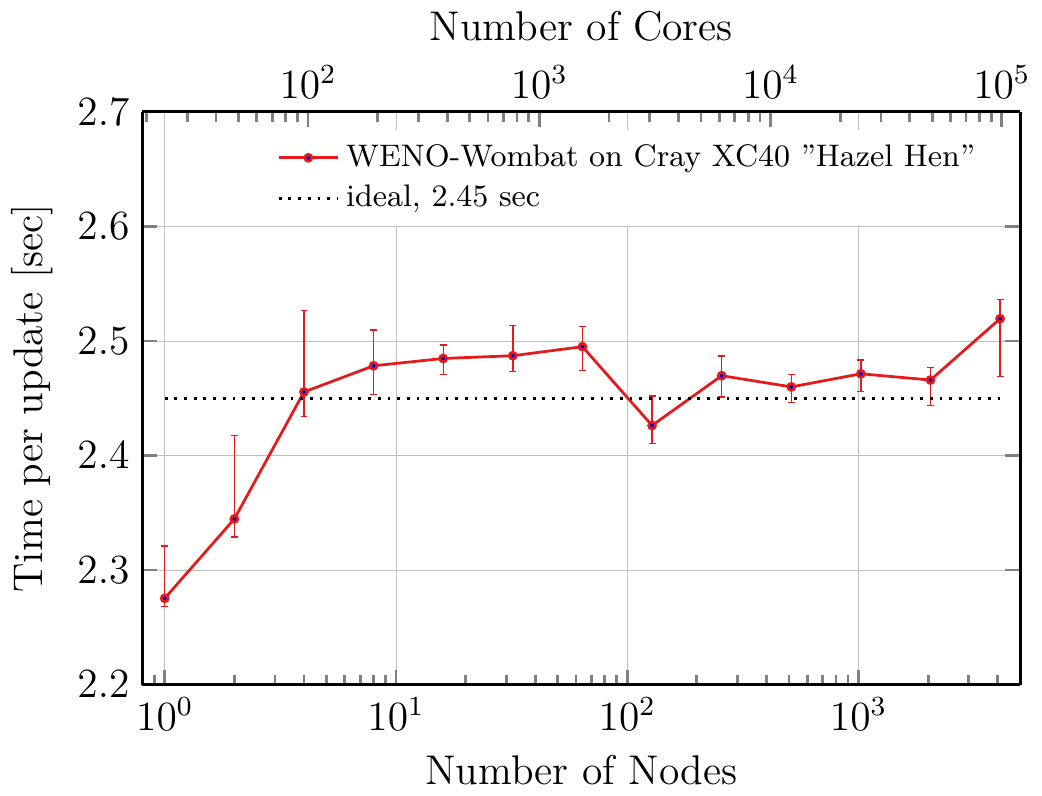}
\caption{Left: Performance in Millions of zones per second per node for different patch sizes on 27 nodes with Intel Broadwell CPUs for a problem of approximate $1024^3$ zones. Right: Weak scaling on the Cray XC40 ''Hazel Hen'' at HLRS Stuttgart, Germany. Shown is the mean time per update of 100 updates over the number of nodes/cores. Error bars indicate the largest and smallest timestep of 100, variations are due to node variablity of the shared system.}
\label{fig.perf}
\end{figure}

On a single Broadwell core, the WENO5 implementation is found to perform at about 4.9 GFlops or 20\% of double precision peak, due to the high degree of SIMD vectorization of the code. On a single Broadwell node, the code performs at 7.2\% of peak, or 1.3 GFlops per core.\par
We showcase the performance of the implementation on multiple nodes, by running a cache blocking study on a Cray Inc. development system. To this end, we vary the patch size of a problem at roughly $1024^3$ zones on 27 nodes with Aries interconnect and 2x18 core Intel Broadwell CPUs per node. We vary the patch size between $4^3$ and $72^3$ zones and the number of MPI ranks per node between 2 and 36, which corresponds to 18 and 1 OpenMP threads. The resulting throughput in Million zones per second per node is shown in figure \ref{fig.perf} left. The optimal patch size is found to be about $18^3$ zones, performing at 0.6 Million zones per second per node. At this patch size, most of the patch fits the level 3 cache of the Broadwell CPU,but not the level 2 cache. We note that \wombat's TVD+CTU scheme has an optimal patch size of $32^3$ on the same system and a strong drop in throughput towards larger patch sizes. Here our implementation approaches optimal performance again. We speculate that because TVD+CTU does not implement array flattening in the algorithm, the hardware pre-fetching is not efficiently keeping data in the cache hierarchy, leading to the drop in performance. In contrast, the flattened array in the WENO5 implementation lead to stride 1 array accesses throughout the algorithm, which  allows the pre-fetcher to keep data in the cache hierarchy efficiently for large patch sizes. Nonetheless, smaller patches are highly desirable for load balancing and thus $16^3-20^3$ is the optimal patch size for WENO5 applications.\par
In weak scaling tests, the problem size is increased alongside the machine size to expose the degree of non-parallelizeable overhead in a program \citep{Amdahl:1967:VSP:1465482.1465560}. We chose a small problem size with a runtime of 2.4 sec per step to clearly expose communication overhead at scale. On the right of figure \ref{fig.perf}, we show the time per update / weak scaling of the WENO5 implementation on the Cray XC40 supercomputer ''Hazel Hen'' at the HLRS in Stuttgart.  The system was \emph{not} dedicated to the test, thus the interconnect is subject to the usual contention seen on large shared systems in practice. From 4 nodes / 96 cores onwards, the step time is found relatively constant between 2.4 and 2.5 seconds per step with a 10\% scatter among time steps, represented by the error bars. This scatter is typical for Haswell system of this size. We note that smallest data point corresponds to workstation size machine with 48 cores. The largest run used 4096 nodes/ 96.000 cores, which is  more than half of the machine. With a problem size of 4.5 Billion zones such a run would resolve the Alfv\'{e}n scale in a triple zoomed cosmological simulation of a galaxy cluster.  
 
\section{Conclusions}

We have presented a new implementation of a fifth order WENO5 scheme for constrained transport magneto-hydrodynamics in the \wombat framework. Our code is aimed at the simulation of cosmic magnetic fields in galaxy clusters, in particular the turbulent small-scale dynamo in the ICM and its Faraday tomography signal. We have motivated the need for new community codes for this particular problem as supercomputers enter the exasflops regime. We have given a concise overview of the WENO5 algorithm and the implementation in \wombat. Finally we shown a few code tests with the new code and argued that the algorithm represents an efficiency optimum as it doubles the effective resolution compared to second order codes common in the field today. We have also shown that given the same resolution, WENO5 resolves instabilities better  than second order TVD+CTU and improves on magnetic energy conservation. Finally we have shown cache optimization tests and demonstrated excellent weak scaling of the code up to realistic problem sizes of 4.5 billion zones on a current Cray  XC40 supercomputer. Thus we are confident that accurate predictions of the magnetic field distribution in galaxy clusters from the small scale dynamo with resolved Alfv\'{e}n scale are within reach in the next 2 years.

\acknowledgments{We would like to thank the two referees for constructive criticism that improved the paper significantly. This research has received funding from the People Programme (Marie Sklodowska Curie Actions) of the European Unions Eighth Framework Programme H2020 under REA grant agreement no 658912, ”Cosmo Plasmas”. Access to the ’Hazel Hen’ at HLRS has been granted through PRACE preparatory access project ”PRACE 4477”. All graphs in the work have been produced using PGF and using the Julia\footnote{\url{julialang.org}} programming language. TWJ acknowledges support from the US NSF through grant AST1714205 }



\reftitle{References}
\externalbibliography{yes}
\bibliography{master}

\end{document}